\PassOptionsToPackage{table}{xcolor}  
\documentclass{article}
\usepackage{cmap}     
\usepackage[T1]{fontenc}
\usepackage{tipa}
\usepackage{extraipa}  
\usepackage[utf8]{inputenc}
\usepackage{_etc/ismir} 
\usepackage{amsmath,cite,url}
\usepackage{bm}   
\usepackage{graphicx}
\usepackage{color}
\usepackage{tcolorbox}
\usepackage{xcolor}
\usepackage{array}     
\usepackage{booktabs}
\usepackage{multirow}

\definecolor{preserved}{HTML}{1D9E75}
\definecolor{attenuated}{HTML}{BA7517}
\definecolor{absent}{HTML}{A32D2D}
\definecolor{novel}{HTML}{666666}
 
\definecolor{presbg}{HTML}{2F6DA4}   \definecolor{presink}{HTML}{FFFFFF}
\definecolor{attbg}{HTML}{C1D3E4}    \definecolor{attink}{HTML}{15537E}
\definecolor{absbg}{HTML}{F5F8FB}    \definecolor{absink}{HTML}{6B6A66}
\definecolor{novbg}{HTML}{FFFFFF}    \definecolor{novink}{HTML}{E0A106}

\newcommand{\vnov}{\textcolor{novink}{\emph{novel}}}

\definecolor{revbg}{HTML}{CF8C89}    \definecolor{revink}{HTML}{8B2320}

\newcommand{\vkey}[3]{{\setlength{\fboxsep}{0pt}%
  \colorbox{#1}{\textcolor{#2}{\,\strut\emph{#3}\,}}}}
\newlength{\vkeywd}
\definecolor{reversedc}{HTML}{6A4C93}

\title{Vocal Music under Phoneme-Conditional Analysis}

\multauthor
  {Hayoon Kim$^{\flat\natural}$ \hspace{1cm} Kyogu Lee$^{\flat\natural\sharp\dagger}$}
  {
      $^\flat$Music and Audio Research Group, Seoul National University\\
      $^\natural$Department of Intelligence and Information, Seoul National University\\
      $^\sharp$AIIS, Seoul National University \quad $^\dagger$IPAI, Seoul National University\\
      {\tt\small \{hyway, kglee\}@snu.ac.kr}
  }
\def\authorname{H. Kim and K. Lee}

\usepackage[bookmarks=false,hidelinks,breaklinks=true]{hyperref}
\hypersetup{pdftitle={Vocal Music under Phoneme-Conditional Analysis},
            pdfauthor={Hayoon Kim, Kyogu Lee},
            pdfsubject={Proceedings of the 27th International Society for
                        Music Information Retrieval Conference},
            pdfkeywords={singing voice, phonetics, forced alignment,
                         cross-linguistic, music information retrieval},
            pdflang={en}}

\begin{document}
\maketitle

\begin{abstract} 
The vocal music of each language carries a distinctive sonic identity, even without instrumental accompaniment. We ask whether these differences are measurable and traceable to specific phonemes. To tackle this question, we introduce phoneme-conditional analysis, which isolates the acoustic effect of typologically distinctive phonemes by comparing marker syllables against matched non-marker controls within the same song, holding singer, melody, and genre constant. Across nine typologically diverse languages and thousands of songs, we measure effects along five acoustic dimensions. Song-level profiles built from these effects identify the language of an unaccompanied vocal at 85.5\% balanced accuracy in a nine-way classification with folds grouped by artist; whether the separability arises by accumulation of the phoneme-local effects themselves is left open. Our findings suggest that phonological structure leaves systematic and measurable traces in how each language is sung.
\end{abstract}

\section{Introduction}\label{sec:introduction}

\begin{figure*}
    \centering
    \begin{tcolorbox}[colback=gray!10, colframe=black, title=Three-Level Hypothesis]

    \colorbox{black}{\textcolor{white}{H1}} Phoneme-level articulatory and acoustic properties directly shape the musical realization of their host syllables

    \colorbox{black}{\textcolor{white}{H2}} Recurrent H1 effects accumulate to form language-specific patterns in vocal music

    \colorbox{black}{\textcolor{white}{H3}} These patterns become conventionalized and persist as part of a musical tradition's identity
    
    \end{tcolorbox}
    \caption{Three-level hypothesis linking phoneme-level constraints to language-specific musical patterns and their conventionalization in musical traditions.}
    \label{fig:hypothesis}
\end{figure*}

Arabic popular music and Japanese pop sound fundamentally different, not merely in language but in how pitch, timing, and vocal quality themselves behave. Arabic singing favors dense micro-ornamentation around a narrow melodic core and flexes syllable durations around the beat \cite{racy2004making}; Japanese pop distributes syllables with near-metronomic uniformity and avoids melisma \cite{tokita2008ashgate}.
Neither pattern is genre-specific, since Arabic ornamentation persists from maqam to contemporary pop \cite{frishkopf2010music} and Japanese syllabic evenness from enka to J-pop \cite{wade2005music}---suggesting these patterns reflect the languages themselves, not musical convention alone.

The hypothesis that linguistic structures shape musical patterns is supported by a robust body of empirical evidence. Rhythmic variability in instrumental music mirrors the prosodic timing of the composer's native language \cite{patel2003empirical}; lexical tone constrains melodic contour, near-obligatorily in Cantonese pop and more loosely in Mandarin \cite{wong2002can}; and text-setting maps linguistic stress to musical meter along language-specific phonological lines \cite{hayes2009textsetting}.

Despite these insights, critical gaps limit the explanatory power of the existing literature. 
First, rhythm studies have relied on symbolic notation \cite{temperley2017rhythmic}, where speech--music correlations often weaken or reverse; sub-segmental durations that survive quantization to the metrical grid go unmeasured (Resolution gap).
Second, tone-melody research has established how pitch-bearing features shape melody, yet consonant inventories' influence on singing remains virtually unexamined (Phonological gap).
Third, rhythm, tone, and timbre are typically studied in isolation on small samples \cite{savage2015statistical}, leaving no unified synthesis of a language's vocal identity (Integration gap).

To bridge these gaps, we propose three hypotheses (H1--H3) spanning sub-segmental mechanics to systemic organization as elaborated in Figure~\ref{fig:hypothesis}. We rigorously test H1, provide preliminary evidence for H2, and defer H3 to future work.

H1 (Phoneme-local constraints) posits that the articulatory and acoustic properties of individual speech sounds physically affect the musical realization of their host syllables: a pharyngeal constriction alters the formant structure of adjacent vowels, a complex onset cluster delays the syllable's perceptual center, a moraic nasal demands its own rhythmic slot. Being consequences of vocal tract physics during singing, these effects are isolable through within-song matched controls. H2 (Emergent conventions) holds that sufficiently frequent H1 effects come to characterize a language's vocal music at large---its default rhythmic profile, ornamentation strategies, and characteristic vocal quality. H3 (Tradition autonomy) proposes that such conventions persist beyond the phonological contexts that produced them, as when a singer's native-language habits transfer to second-language performance.

Guided by this framework, our contributions are threefold: (1) a three-level framework separating phoneme-local mechanics from language-wide convention, which attributes the weak notation-level rhythm correlations to measurement resolution; (2) a phoneme-conditional analysis method using within-song matched controls; 
and (3) application to 9 languages demonstrating language-specific acoustic signatures, enabling $85.5\%$ nine-way classification accuracy with artist-grouped folds.\footnote{\raggedright Code and materials are available at \url{https://github.com/gillosae/phoneme-conditional-singing-analysis}.}
\section{Related Works}\label{sec:related_works}
\begin{table*}[t]
  \centering
  \small
  \setlength{\tabcolsep}{3.5pt}
  \begin{tabular}{@{}llclllll@{}}
    \toprule
    \textbf{Language} & \textbf{Family} & \textbf{Type} & \textbf{Primary marker} & \textbf{Rarity} & \textbf{Acoustic cue} & \textbf{G2P} & \textbf{Charts} \\
    \midrule
    Arabic (Eg.)
      & Afroasiatic & A
      & Pharyngeal /\textrevglotstop~\textcrh/, emphatic /t\textsuperscript{\textrevglotstop}~s\textsuperscript{\textrevglotstop}/
      & {$<$5\%}
      & F1$\uparrow$, F2$\downarrow$
      & \texttt{epitran}
      & EG \\
    English
      & IE--Germanic & A
      & Schwa /\textipa{@}/, approximant /\textturnr/
      & {$<$5\%}
      & F3 lowering
      & \texttt{eSpeak}
      & US, UK \\
    French
      & IE--Romance & A
      & Nasal V /\textipa{\~E \~O \~A}/, uvular /\textinvscr/
      & {$\sim$25\%}
      & A1--P0, F2
      & \texttt{epitran}
      & FR \\
    Hindi
      & IE--Indo-Aryan & A
      & Retroflex /\textrtailt~\textrtaild/, breathy /b\textsuperscript{H}~d\textsuperscript{H}/
      & {$<$5\%} 
      & F3 trans., H1--H2
      & \texttt{epitran}
      & IN \\
    Japanese
      & Japonic & A
      & Moraic geminates /k\textlengthmark~t\textlengthmark/, nasal /\textscn/, bilabial /\textphi/
      & {$<$5\%}
      & Closure dur.\
      & \texttt{pykakasi}
      & JP \\
    Korean
      & Koreanic & A
      & 3-way laryngeal (asp/tense/lax)
      & {$<$1\%}
      & VOT, H1--H2
      & \texttt{g2pK2}
      & KR \\
    Turkish
      & Turkic & A
      & High back unrounded /\textturnm/
      & {$\sim$6\%}
      & F1$\downarrow$, F2$\downarrow$
      & \texttt{epitran}
      & TR \\
    Mandarin
      & Sino-Tibetan & A+T
      & Retroflex /\textrtailt\textrtails~\textrtails~\textturnrrtail/ + 4 tones
      & {$\sim$6\%}
      & Spec.\ COG, F0
      & \texttt{pypinyin}
      & TW \\
    Swedish
      & IE--Germanic\ & A+T
      & Sje-sound /\texththeng/ + long V, pitch accent
      & {$<$1\%}
      & Spec.\ COG, F0
      & \texttt{eSpeak}
      & SE \\
    \bottomrule
  \end{tabular}
  \caption{Language selection and phonological markers.
  \emph{Type~A}: articulatory markers measured via within-song matched comparison.
  \emph{A+T}: articulatory plus tonal or pitch-accent contrasts that may interact with melodic pitch.
  \emph{G2P}: grapheme-to-phoneme tool; all output is IPA, Unicode NFC-normalized.
  \emph{Charts}: Spotify daily charts by country.
  \emph{Rarity}: approximate share of the world's languages with the feature; where a marker bundles several segments, the rarest component is given.}
  \label{tab:lang_selection}
\end{table*}
\subsection{Rhythmic Variations and the Vocal Paradox}
The normalized Pairwise Variability Index (nPVI)~\cite{grabe02durational} is the standard metric for comparing linguistic and musical rhythm. Following the English--French comparison of instrumental themes~\cite{patel2003empirical}, the stress-timed/syllable-timed mirroring has been replicated across languages~\cite{sadakata2004cross}, genres~\cite{vanhandel2010role}, and periods~\cite{temperley2011music}, but in vocal music these differences diminish or reverse~\cite{temperley2017rhythmic}---weakest where the language is actually sung. Notated durations take few distinct values, so nPVI over them largely counts adjacent notes written equal~\cite{condit2019deconstructing}: symbolic notation, rather than the speech--music link itself, is the limiting factor (Resolution gap). We therefore compute nPVI over forced-aligned phoneme durations.

\subsection{Tone--Melody Correspondence and its Boundaries}
Lexical tone constrains melodic contour to a tradition-dependent degree, from over 90\% parallel tone--melody motion in Cantonese pop~\cite{wong2002can} to markedly looser alignment elsewhere~\cite{kirby2016tone}. Methodologically this literature converges on the bigram level: what matters is whether successive tone and pitch movements are parallel or contrary~\cite{mcpherson2018tone, ladd2020tone}. Text-setting work extends this pitch-centric paradigm to prosodic weight and to syllable perceptual centers, the instant a syllable is heard to begin, which its onset consonants displace~\cite{ryan2022syllable,hayes1996role}; but the acoustic consequences of segmental inventories---pharyngeals, ejectives, retroflexes---remain unaddressed (Phonological gap).

\subsection{Phonetic Baselines and Multilingual Corpora}
Marker selection rests on phonemes with well-characterized acoustic signatures~\cite{ladefoged1996sounds}: F3 lowering for English /\textturnr/~\cite{hagiwara1995acoustic}, F2 lowering for Arabic emphatics~\cite{jongman2011acoustics}, VOT and f0-onset separation in the Korean laryngeal contrast~\cite{cho1999variation, cho2002acoustic}, lengthening of French nasal vowels~\cite{delattre1968role}, and closure-duration ratios for Japanese geminates~\cite{han1994acoustic,kawahara2015phonetics}. These are the denominators of our preservation ratios (Table~\ref{tab:preservation}). On the corpus side, multilingual singing datasets~\cite{zhang2024gtsinger} and anthropological surveys~\cite{savage2015statistical, mehr2019universality} establish cross-cultural regularities at the level of whole utterances or pieces; songs are globally slower, higher, and more pitch-stable than speech~\cite{ozaki2024globally}. What remains open is the mechanism---which phonological properties produce these differences, and how phoneme-local effects aggregate into a tradition's characteristic sound (Integration gap), which our within-song design targets. 
\section{Method}\label{sec:method}

\subsection{Language Selection}
We select nine languages (Table~\ref{tab:lang_selection}) spanning six families and three prosodic types (stress-, syllable-, and mora-timed), maximizing typological diversity subject to chart-data availability. Each contributes one or more \textit{markers}: phonemes typologically rare ($<$20\% of the world's languages per WALS/PHOIBLE~\cite{wals,phoible}) and acoustically measurable from audio, the one exception being French nasal vowels at ${\sim}25\%$, admitted for the sake of a Romance language. Seven languages carry pure Type~A markers, whose articulatory properties constrain the sung syllable; Mandarin and Swedish add Type~T (tonal or pitch-accent) markers, testing whether F0-based contrasts survive melodic constraint. For Arabic we use the Egyptian dialect~\cite{danielson2008voice}.

\subsection{Corpus Construction}
For each language we sample 1{,}000--2{,}100 songs from Spotify national daily charts~\cite{spotify_charts} via a public aggregator~\cite{kworb_spotify}, matching metadata to YouTube audio-only streams at 48\,kHz / 24-bit FLAC. Language identification accepts a track when at least two of four signals agree---GlotLID on title/description, a channel allowlist, MMS-LID-4017 on audio, and GlotLID on Whisper transcripts---and duplicates are removed by Chromaprint fingerprinting. After separation, transcription, and alignment, 9{,}331 tracks carry a phone-level alignment, of which 5{,}024 pass the alignment-agreement gate of Section~\ref{sec:pipeline} (370--751 per language). All analyses use this gated set; we use only metadata and feature vectors.

\subsection{Processing Pipeline}\label{sec:pipeline}

\noindent\textbf{Source separation ---}
We isolate vocals using Mel-Band RoFormer~\cite{wang2024mel}, retaining only tracks with an estimated SDR greater than 3\,dB.

\noindent\textbf{Lyrics and G2P ---}
Lyrics come from pre-collected databases~\cite{lrclib, musixmatch, genius}; grapheme-to-phoneme conversion uses language-specific tools~\cite{espeakng, g2pk2, mortensen-etal-2018-epitran, pykakasi, pypinyin} (Table~\ref{tab:lang_selection}).

\noindent\textbf{Forced alignment ---}
Each track is aligned in a single CTC pass over the whole song, avoiding dependence on ASR word timestamps. Two acoustic models run in parallel and are averaged: MMS Forced Alignment~\cite{pratap2024scaling} over uroman transliteration~\cite{hermjakob2018out}, and a wav2vec2 CTC model emitting espeak IPA symbols directly, fine-tuned on 5{,}062 sung utterances from GTSinger~\cite{zhang2024gtsinger}. Their errors are nearly independent ($r = 0.09$ on NUS-48E) and averaging helps: the blend places 74.6\% of onsets within 50\,ms against 72.3\% and 69.5\% for the two models. Blank frames between CTC token spans are redistributed to neighboring phones, and each onset is snapped to the strongest spectral change within $\pm$30\,ms when locally prominent. Because both models see the same phone sequence, their median per-track disagreement serves as a ground-truth-free confidence signal; tracks disagreeing by more than 200\,ms are excluded (alignment-agreement gate). The gate selects on transcript completeness rather than on performer: excluded tracks carry a median of 402 lyric characters against 1{,}678 for retained ones, and 723 of the 1{,}914 retained artists also appear among the excluded.

\noindent\textbf{F0 extraction ---}
Fundamental frequency comes from an ensemble of FCPE~\cite{luo2025fcpe} and RMVPE~\cite{wei2023rmvpe}. Frames below 0.3 RMVPE voicing confidence are masked; frames where the estimators disagree by $>$2 semitones receive halved confidence and are re-thresholded, filtering instrumental leakage from imperfect separation.

\noindent\textbf{Feature extraction ---}
For each aligned phoneme we extract features from the central 70\% of the segment, in five modules: \textit{articulatory impact} (F1--F3, spectral tilt, H1--H2, HNR, and F1/F2 deltas to neighbouring phonemes), \textit{melodic deviation} (F0 mean, range, slope, sd), \textit{rhythmic adaptation} (duration), \textit{melisma} (notes per phoneme), and \textit{timbral signature} (spectral centroid, MFCC). Text-setting alignment applies only to Mandarin and Swedish. Song-level aggregates are nPVI and melisma rate.

\begin{table*}[t]
  \centering
  \footnotesize
  \setlength{\tabcolsep}{2.5pt}
  \renewcommand{\arraystretch}{1.25}
  \providecommand{\colhead}[1]{\shortstack{#1}}
  \providecommand{\colheadc}[1]{\raisebox{2.8pt}[0pt][0pt]{#1}}
  \begin{tabular}{@{}l@{\hspace{10pt}}l@{\hspace{10pt}}>{\centering\arraybackslash}p{23pt}>{\centering\arraybackslash}p{23pt}>{\centering\arraybackslash}p{23pt}>{\centering\arraybackslash}p{23pt}>{\centering\arraybackslash}p{23pt}>{\centering\arraybackslash}p{23pt}>{\centering\arraybackslash}p{23pt}>{\centering\arraybackslash}p{23pt}>{\centering\arraybackslash}p{23pt}>{\centering\arraybackslash}p{23pt}>{\centering\arraybackslash}p{23pt}>{\centering\arraybackslash}p{23pt}@{\hspace{5pt}}>{\centering\arraybackslash}p{24pt}@{}}
    \toprule
    \noalign{\vskip-2pt}
    \multirow{2}{*}{\textbf{Category}} & \multirow{2}{*}{\textbf{Examples}} & \multicolumn{6}{c}{Transition} & \multicolumn{3}{c}{$\Delta$F} & \multicolumn{3}{c}{F0} & \multirow{2}{*}{$\bm{|\overline{d}|}$} \\
    \noalign{\vskip-3pt}
    \cmidrule(lr){3-8}\cmidrule(lr){9-11}\cmidrule(lr){12-14}
    \noalign{\vskip-3pt}
    & & F1 & F2 & F3 & tilt & HNR & \makebox[0pt][c]{H1--H2} & F1 & F2 & F3 & dev & range & slope & \\
    \noalign{\vskip-2pt}
    \midrule
    Retroflex & /\textrtailt~\textrtails/ & \cellcolor[HTML]{C6736E}$+0.16$ & \cellcolor[HTML]{D28F8C}$+0.12$ & \cellcolor[HTML]{B0403A}\textcolor{white}{$+0.23$} & \cellcolor[HTML]{B03F39}\textcolor{white}{$+0.23$} & \cellcolor[HTML]{1F4E79}\textcolor{white}{$-0.25$} & \cellcolor[HTML]{F0DEDC}$+0.04$ & \cellcolor[HTML]{D18D8A}$+0.12$ & \cellcolor[HTML]{C0645F}$+0.18$ & \cellcolor[HTML]{D3918E}$+0.12$ & \cellcolor[HTML]{EFD8D6}$+0.04$ & \cellcolor[HTML]{EBCAC8}$+0.06$ & \cellcolor[HTML]{B3CADE}$-0.08$ & 0.14 \\
    High back unrounded & /\textturnm/ & \cellcolor[HTML]{ADC5DC}$-0.08$ & \cellcolor[HTML]{DDA9A6}$+0.10$ & \cellcolor[HTML]{C2D3E4}$-0.06$ & \cellcolor[HTML]{AD3933}\textcolor{white}{$+0.24$} & \cellcolor[HTML]{BCCFE2}$-0.07$ & \cellcolor[HTML]{C77671}$+0.15$ & \cellcolor{white}{---} & \cellcolor{white}{---} & \cellcolor{white}{---} & \cellcolor[HTML]{F6F2F1}$+0.01$ & \cellcolor[HTML]{DBE4ED}$-0.03$ & \cellcolor[HTML]{F6F3F2}$+0.00$ & 0.08 \\
    Tense & /\textipa{\subdoublevert{p}~\subdoublevert{t}}/ & \cellcolor[HTML]{CAD9E7}$-0.05$ & \cellcolor[HTML]{EAC6C4}$+0.07$ & \cellcolor[HTML]{EFD9D8}$+0.04$ & \cellcolor[HTML]{DDA9A6}$+0.10$ & \cellcolor[HTML]{769FC4}$-0.13$ & \cellcolor[HTML]{EFD8D6}$+0.04$ & \cellcolor[HTML]{E4EAF0}$-0.02$ & \cellcolor[HTML]{EDD1CF}$+0.05$ & \cellcolor[HTML]{F1DFDE}$+0.03$ & \cellcolor[HTML]{CD8480}$+0.14$ & \cellcolor[HTML]{F5F1EF}$+0.01$ & \cellcolor[HTML]{F0DBD9}$+0.04$ & 0.06 \\
    Bilabial fricative & /\textphi~\textbeta/ & \cellcolor[HTML]{F6F2F1}$+0.01$ & \cellcolor[HTML]{265B8B}\textcolor{white}{$-0.23$} & \cellcolor[HTML]{F3E6E5}$+0.02$ & \cellcolor[HTML]{A9C3DA}$-0.09$ & \cellcolor[HTML]{F3F4F5}$-0.00$ & \cellcolor[HTML]{EDD1CF}$+0.05$ & \cellcolor[HTML]{F6F6F6}$-0.00$ & \cellcolor[HTML]{B5CBDF}$-0.07$ & \cellcolor[HTML]{F5EEEC}$+0.01$ & \cellcolor[HTML]{5F8FBA}$-0.15$ & \cellcolor[HTML]{E9C4C2}$+0.07$ & \cellcolor[HTML]{F5EFEE}$+0.01$ & 0.06 \\
    Aspirated & /p\textsuperscript{h}~k\textsuperscript{h}/ & \cellcolor[HTML]{F0DCDB}$+0.04$ & \cellcolor[HTML]{EAC8C6}$+0.06$ & \cellcolor[HTML]{F3E6E5}$+0.02$ & \cellcolor[HTML]{F0DBD9}$+0.04$ & \cellcolor[HTML]{2B6599}\textcolor{white}{$-0.20$} & \cellcolor[HTML]{CE8581}$+0.13$ & \cellcolor[HTML]{F3E8E7}$+0.02$ & \cellcolor[HTML]{F3E9E8}$+0.02$ & \cellcolor[HTML]{F3E8E7}$+0.02$ & \cellcolor[HTML]{F5EEEC}$+0.01$ & \cellcolor[HTML]{ECCCCB}$+0.06$ & \cellcolor[HTML]{F4F5F5}$-0.00$ & 0.05 \\
    Nasal vowel & /\textipa{\~E \~O \~A}/ & \cellcolor[HTML]{F2E2E1}$+0.03$ & \cellcolor[HTML]{DDE5ED}$-0.03$ & \cellcolor[HTML]{E6BEBC}$+0.07$ & \cellcolor[HTML]{E4EAF0}$-0.02$ & \cellcolor[HTML]{CC827E}$+0.14$ & \cellcolor[HTML]{E5BAB8}$+0.08$ & \cellcolor{white}{---} & \cellcolor{white}{---} & \cellcolor{white}{---} & \cellcolor[HTML]{F4EBE9}$+0.02$ & \cellcolor[HTML]{F3E6E5}$+0.02$ & \cellcolor[HTML]{F4EBE9}$+0.02$ & 0.05 \\
    Schwa/rhotic & /\textipa{@}~\textturnr~\textrhookschwa/ & \cellcolor[HTML]{E4EAF0}$-0.02$ & \cellcolor[HTML]{EDD2D0}$+0.05$ & \cellcolor[HTML]{EDD2D0}$+0.05$ & \cellcolor[HTML]{C97B76}$+0.15$ & \cellcolor[HTML]{9FBCD5}$-0.10$ & \cellcolor[HTML]{EDD2D0}$+0.05$ & \cellcolor[HTML]{F6F5F4}$+0.00$ & \cellcolor[HTML]{F4F5F5}$-0.00$ & \cellcolor[HTML]{EAEFF2}$-0.02$ & \cellcolor[HTML]{B9CEE1}$-0.07$ & \cellcolor[HTML]{E0E8EE}$-0.03$ & \cellcolor[HTML]{DDE5ED}$-0.03$ & 0.05 \\
    Pharyngeal/emphatic & /\textrevglotstop~\textcrh~t\textsuperscript{\textrevglotstop}/ & \cellcolor[HTML]{DFADAA}$+0.09$ & \cellcolor[HTML]{B9CEE1}$-0.07$ & \cellcolor[HTML]{F6F2F1}$+0.01$ & \cellcolor[HTML]{D8E2EC}$-0.04$ & \cellcolor[HTML]{C5D6E5}$-0.06$ & \cellcolor[HTML]{E4EAF0}$-0.02$ & \cellcolor[HTML]{F3E9E8}$+0.02$ & \cellcolor[HTML]{E7ECF1}$-0.02$ & \cellcolor[HTML]{F6F2F1}$+0.01$ & \cellcolor[HTML]{C2D3E4}$-0.06$ & \cellcolor[HTML]{ECCBC9}$+0.06$ & \cellcolor[HTML]{F6F6F6}$-0.00$ & 0.04 \\
    Uvular & /q~\textchi~\textinvscr/ & \cellcolor[HTML]{F4EBE9}$+0.02$ & \cellcolor[HTML]{ECF0F2}$-0.01$ & \cellcolor[HTML]{F6F2F1}$+0.01$ & \cellcolor[HTML]{F6F6F6}$-0.00$ & \cellcolor[HTML]{E0E8EE}$-0.03$ & \cellcolor[HTML]{C0D2E3}$-0.07$ & \cellcolor[HTML]{F6F6F6}$-0.00$ & \cellcolor[HTML]{E9EDF1}$-0.02$ & \cellcolor[HTML]{ECF0F2}$-0.01$ & \cellcolor[HTML]{F3E9E8}$+0.02$ & \cellcolor[HTML]{F5EEEC}$+0.01$ & \cellcolor[HTML]{E9EDF1}$-0.02$ & 0.02 \\
    \bottomrule
  \end{tabular}
  \caption{Paired Cohen's $d$ for vowels adjacent to marked versus control
  consonants. \textemdash{} marks cells with $<$30 pairs.}
  \label{tab:phoneme_property}
\end{table*}

\subsection{Alignment Validation}\label{sec:validation}
On NUS-48E~\cite{duan2013nus} (47 sung recordings with manual phone boundaries) our aligner places onsets with a median error of 23\,ms, 74.6\% within 50\,ms, and recovers durations at a median $0.85$ of true length. The underestimate is compressive rather than constant, inflating phones under 81\,ms (pred/true $1.38$) and cutting those over 210\,ms ($0.53$). Two consequences follow for Section~\ref{sec:preservation}. Random boundary error inflates variance without biasing a paired difference: with the per-phone error centred at $-18$\,ms and $\mathrm{sd} = 148$\,ms---a conservative figure, set by a tail of gross misplacements---it adds $0.9$--$1.7$\,ms of standard error over the pairs behind the three duration effects. Compression attenuates: a long ($\geq$150\,ms) versus short ($\leq$100\,ms) pairing of the same class recovers $34\%$ of the true difference ($+78$ against $+228$\,ms), so our millisecond magnitudes are lower bounds; being monotone in true duration it shrinks a contrast without reversing its sign, though this does not cover class-specific bias. As a floor on the aligner's own contribution, phones paired by class and true-duration bin---true difference $+0.06$\,ms---measure $-1.6$\,ms, 95\% CI $[-4.0, +0.6]$; this is English sung material, the only corpus here with manual boundaries, assumed to transfer. Two alternatives do markedly worse. Distributing lyric words across ASR word intervals by index gives 443\,ms median onset error, 14.6\% within 50\,ms, and a $0.18$ duration ratio. Given the same phone sequences, the Montreal Forced Aligner~\cite{mcauliffe2017montreal} aligned only 8 of 47 sung recordings usably; on spoken recordings by the same singers---a different subset---it reaches 20.0\,ms against our 22.9\,ms.

\subsection{Marker Classification and Matched Pairing}\label{sec:matching}
Each phoneme is labeled marker or non-marker by language-specific rules from the phonological typology literature (Table~\ref{tab:lang_selection}). For each marker we select a matched control in the same song by additive scoring: phone-type match (weight 10; vowel markers additionally matched on height$\times$backness) with temporal proximity ($1/(1+|i-j|)$) as tiebreaker, each control reused at most three times. Korean is further decomposed into aspirated--lax, tense--lax, and aspirated--tense comparisons; Type T markers, where every syllable bears a tone, instead pair contour tones against level tones matched on rime; Swedish pitch accent yields too few such pairs to fit.

\subsection{Statistical Analysis}\label{sec:stats}

\noindent\textbf{Within-song effect sizes ---}
For each language $\times$ dimension cell we compute paired Cohen's $d$ between marker and matched control, taking $\max|d|$ across the features in that dimension, so the reported magnitude is an upper bound on the dimension's effect. Significance uses permutation tests (5{,}000 iterations) with the null built on the same maximum statistic, Bonferroni-corrected across the 50 cells ($\alpha = 0.05/50$). Because each cell rests on $3{,}855$--$153{,}849$ pairs, significance is near-automatic; we therefore fix a practical floor of $|d| \geq 0.1$ in advance and read cells below it as significant but negligible.

\noindent\textbf{Linear mixed models ---}
Per-language LMMs of the form \texttt{feature} $\sim$ \texttt{is\_marker} $+$ \texttt{beat\_phase} $+$ $(1|\texttt{song})$, fitted by REML, control for metrical position; where they fail to converge we fall back to Wilcoxon signed-rank tests.

\noindent\textbf{Language classification ---}
Song-level feature vectors (35-dimensional: 20 marker--control difference statistics $+$ 15 raw phoneme distribution statistics) are classified by a Random Forest (200 trees) with 5-fold cross-validation grouped by artist, balanced by undersampling to the smallest class. A song enters only if all 35 features are defined, costing 0--153 tracks per language (Hindi $442 \rightarrow 289$) and leaving 289 songs per language over 1{,}323 artists. Artists recur in chart data---2.0 songs per artist across those 2{,}601 songs---so a song-level split would let the model reach a language through its performers; grouping removes that route at a cost of 1.2 accuracy points. Splitting the vector into its two blocks separates what the marker contrasts contribute from what the raw distributions do: the 20 difference statistics alone reach $79.4\%$ and the 15 distribution statistics alone $55.7\%$, so the distributions are worth $6.1$ points of the $85.5\%$.
\section{Results}\label{sec:results}

All results are computed on the 5{,}024 gated tracks (Section~\ref{sec:pipeline}). Significant effects appear in at least three acoustic dimensions for 9 of the 10 marker sets, the exception being Hindi. Which channel dominates depends on the metric: timbre and voice quality carry the largest effect sizes, while formant transitions are the best preserved of the channels, and then only for retroflexion (Section~\ref{sec:preservation}).

\subsection{Phoneme Properties Shape Singing Acoustics}

Table~\ref{tab:phoneme_property} presents Cohen's $d$ for vowels adjacent to each consonant category, pooled across languages. It reports individual features of the adjacent vowel, whereas Table~\ref{tab:fingerprint} reports the largest $|d|$ over a dimension, so the two are not on the same scale. The dominant channels are voice quality and spectral tilt rather than formant transitions. Retroflexes show the largest single cell ($d = -0.25$, trans HNR; the same row carries $+0.23$ on trans tilt and F3, so the ranking is by $|d|$, not direction), followed by the Turkish high back unrounded vowel ($+0.24$, trans tilt) and bilabial /\textipa{F}/ ($-0.23$, trans F2). Korean aspirated and tense stops separate by channel rather than by magnitude ($-0.20$ HNR vs.\ $+0.14$ F0 deviation), consistent with their phonation-type distinction. Of the 102 filled cells, 21 clear the $|d| \geq 0.1$ floor, 11 reach $0.15$ and 6 reach $0.20$. These are small effects measured precisely rather than large ones.

\subsection{Language-Specific Fingerprints}
Table~\ref{tab:fingerprint} shows paired Cohen's $d$ across five modules per language, ranked by $|\overline{d}|$. Mandarin's dual markers separate along melisma and timbral magnitude rather than along a temporal/spectral axis. Both sets act most strongly on timbre---retroflex markers (Type~A, $0.25$) carry the largest single effect in the corpus ($d = -0.80$, MFCC-1 over $25{,}899$ pairs), the tonal set (Type~T, $0.20$) less than half that ($-0.36$). What is the tonal set's alone is melisma ($+0.28$ vs.\ $-0.07$); rhythmic lengthening is shared ($+0.14$ vs.\ $+0.11$, $\beta = +24.7$~ms; millisecond effects in this section are LMM coefficients, against the raw paired differences of Table~\ref{tab:preservation}), and melodic deviation distinguishes neither ($+0.07$ vs.\ $+0.06$). The timbral column is an MFCC effect in most rows and a spectral centroid for Korean and Mandarin~(T), which is why the $-0.80$ cell has no counterpart in Table~\ref{tab:phoneme_property}, whose columns carry no cepstral features.

Korean is the strongest consonant-marker language, driven by timbre and melisma, followed by Arabic, Turkish and English, each with a timbral component of the same order. The ranking is largely a timbre ranking: dropping that column reorders every row above Japanese. Six marker sets reach significance in all five modules and Japanese three, with melisma avoidance ($-0.11$) and shortening ($-16.4$~ms). Swedish's tonal markers are too sparse to fit separately, so only Swedish~(A) appears. Hindi reaches the fewest modules and is least powered ($4{,}405$ pairs), so its nulls are uninformative, not negative. Pair counts differ by an order of magnitude, so asterisks are not comparable between rows and the ranking is by $\overline{|d|}$.

Splitting Korean pairs by laryngeal category, aspirated--lax and tense--lax yield nearly identical effects while aspirated--tense differences are minimal: aspirated and tense stops pattern together against lax. This follows the Seoul Korean tonogenetic merger~\cite{kang2014voice}, in which F0 onset has largely replaced VOT as the primary laryngeal cue. Melodic pitch does not fully override that cue---the contrast retains ${\sim}30\%$ of its speech magnitude---so the grouping in singing follows the same F0-based partition as contemporary speech.

\begin{table}[t]
  \centering
  \scriptsize
  \setlength{\tabcolsep}{1pt}
  \renewcommand{\arraystretch}{1.25}
  \ifdefined\dwidth\else\newlength{\dwidth}\fi
  \ifdefined\starwd\else\newlength{\starwd}\fi
  \settowidth{\dwidth}{$-0.00$}
  \settowidth{\starwd}{*}
  \providecommand{\numcell}[2]{%
    \hspace*{-0.8\starwd}\makebox[\dwidth][r]{$#1$}\makebox[0pt][l]{#2}}
  \providecommand{\rothead}[1]{%
    \makebox[0pt][l]{\hspace*{-12pt}\rotatebox{22}{\shortstack[l]{#1}}}}
  \begin{tabular}{@{}l@{\hspace{7pt}}>{\centering\arraybackslash}p{24pt}>{\centering\arraybackslash}p{24pt}>{\centering\arraybackslash}p{24pt}>{\centering\arraybackslash}p{24pt}>{\centering\arraybackslash}p{24pt}>{\centering\arraybackslash}p{24pt}@{\hspace{4pt}}r@{}}
    \toprule
    \textbf{Marker set} & \rothead{Articulatory\\[-2.5pt]Impact} & \rothead{Melodic\\[-2.5pt]Deviation} & \rothead{Rhythmic} & \rothead{Melisma} & \rothead{Timbral} & $\bm{|\overline{d}|}$ & \textbf{Pairs} \\
    \midrule
    Mandarin (A) & \cellcolor[HTML]{C0D2E3}\numcell{-0.20}{*} & \cellcolor[HTML]{F3E6E5}\numcell{+0.07}{*} & \cellcolor[HTML]{F1DFDE}\numcell{+0.11}{*} & \cellcolor[HTML]{E4EAF0}\numcell{-0.07}{*} & \cellcolor[HTML]{1F4E79}\textcolor{white}{\numcell{-0.80}{*}} & 0.25 & 25,899 \\
    Korean & \cellcolor[HTML]{BED1E3}\numcell{-0.21}{*} & \cellcolor[HTML]{F1E1DF}\numcell{+0.09}{*} & \cellcolor[HTML]{E5EBF0}\numcell{-0.07}{*} & \cellcolor[HTML]{ADC5DC}\numcell{-0.26}{*} & \cellcolor[HTML]{C97975}\numcell{+0.47}{*} & 0.22 & 48,492 \\
    Mandarin (T) & \cellcolor[HTML]{CCDAE8}\numcell{-0.16}{*} & \cellcolor[HTML]{F3E9E8}\numcell{+0.06}{*} & \cellcolor[HTML]{EFD7D5}\numcell{+0.14}{*} & \cellcolor[HTML]{E0B0AE}\numcell{+0.28}{*} & \cellcolor[HTML]{8EB0CE}\numcell{-0.36}{*} & 0.20 & 153,849 \\
    Arabic & \cellcolor[HTML]{EDD1CF}\numcell{+0.17}{*} & \cellcolor[HTML]{E7ECF1}\numcell{-0.06}{*} & \cellcolor[HTML]{F5F1EF}\numcell{+0.03}{*} & \cellcolor[HTML]{E9EDF1}\numcell{-0.05}{*} & \cellcolor[HTML]{769FC4}\numcell{-0.42}{*} & 0.15 & 25,498 \\
    Turkish & \cellcolor[HTML]{E6BEBC}\numcell{+0.23}{*} & \cellcolor[HTML]{F4ECEB}\numcell{+0.05}{*} & \cellcolor[HTML]{E7ECF1}\numcell{-0.06}{*} & \cellcolor[HTML]{F7F6F5}\numcell{+0.00}{} & \cellcolor[HTML]{8EB0CE}\numcell{-0.35}{*} & 0.14 & 32,277 \\
    English & \cellcolor[HTML]{F0DEDC}\numcell{+0.11}{*} & \cellcolor[HTML]{E7ECF1}\numcell{-0.06}{*} & \cellcolor[HTML]{B9CEE1}\numcell{-0.22}{*} & \cellcolor[HTML]{E0E8EE}\numcell{-0.08}{*} & \cellcolor[HTML]{EAC6C4}\numcell{+0.21}{*} & 0.14 & 78,968 \\
    French & \cellcolor[HTML]{EED4D2}\numcell{+0.15}{*} & \cellcolor[HTML]{F6F2F1}\numcell{+0.02}{*} & \cellcolor[HTML]{EED4D2}\numcell{+0.15}{*} & \cellcolor[HTML]{F2E5E4}\numcell{+0.08}{*} & \cellcolor[HTML]{EAC8C6}\numcell{+0.20}{*} & 0.12 & 34,020 \\
    Japanese & \cellcolor[HTML]{E7ECF1}\numcell{-0.06}{} & \cellcolor[HTML]{E0E8EE}\numcell{-0.08}{*} & \cellcolor[HTML]{DDE5ED}\numcell{-0.10}{*} & \cellcolor[HTML]{D9E3EC}\numcell{-0.11}{*} & \cellcolor[HTML]{E7ECF1}\numcell{-0.06}{} & 0.08 & 15,213 \\
    Hindi & \cellcolor[HTML]{F2E4E2}\numcell{+0.08}{} & \cellcolor[HTML]{F3E6E5}\numcell{+0.07}{*} & \cellcolor[HTML]{EFF2F3}\numcell{-0.03}{} & \cellcolor[HTML]{F6F5F4}\numcell{+0.01}{} & \cellcolor[HTML]{C0D2E3}\numcell{-0.20}{*} & 0.08 & 4,405 \\
    Swedish (A) & \cellcolor[HTML]{F3E8E7}\numcell{+0.06}{*} & \cellcolor[HTML]{F6F2F1}\numcell{+0.02}{*} & \cellcolor[HTML]{F6F5F4}\numcell{+0.01}{} & \cellcolor[HTML]{F6F3F2}\numcell{+0.01}{} & \cellcolor[HTML]{F0DBD9}\numcell{+0.12}{*} & 0.05 & 65,536 \\
    \bottomrule
  \end{tabular}
  \caption{Paired Cohen's $d$ between markers and matched controls over five dimensions. $^*$: Bonferroni-corrected $\alpha = 0.05/50$. \emph{pairs}: matched pairs per set.}
  \label{tab:fingerprint}
\end{table}

\subsection{Speech-to-Singing Preservation Hierarchy}\label{sec:preservation}
Table~\ref{tab:preservation} compares singing effects against published speech baselines. Every row rests on enough pairs to be significant at $p<.001$, so the argument is about the ratio to the baseline rather than about whether the effect is nonzero. Preservation is channel-dependent and does not follow the effect-size ranking: formant coarticulation is best preserved where the constriction is a tongue-body gesture, while duration is uniformly attenuated and in two cases reverses sign.

\noindent\textbf{Duration attenuated ---}
No durational contrast reaches half its speech baseline. Mandarin tonal lengthening retains ${\sim}30\%$ ($+26$~ms vs.\ ${\sim}50$--$120$~ms~\cite{duanmu2007phonology}) and French nasal vowels ${\sim}19\%$ ($+7$ vs.\ $+40$~ms). Korean aspirated/tense stops reverse: $10$~ms shorter than lax controls in singing, against ${\sim}25$--$40$~ms longer in speech. Japanese geminates remain shortened ($-15$~ms), consistent with absorption into the beat grid. These three are 4--9$\times$ the aligner's null floor ($1.6$~ms, Section~\ref{sec:validation}) against 16$\times$ for Mandarin, and both reversals fall on long consonants, where a boundary inside the closure does most damage; we therefore read them as directions rather than magnitudes.

\noindent\textbf{Timbral shape attenuated ---}
The Mandarin retroflex centroid shift measures $+1{,}342$~Hz, ${\sim}45\%$ of its speech baseline: attenuated, but the largest spectral effect in the corpus.

\noindent\textbf{Formants reduced outside retroflexion ---}
The Mandarin retroflex F2 shift is fully preserved ($+543$ vs.\ ${\sim}500$~Hz) and Hindi retroflex F3 lowering reaches ${\sim}33\%$ ($-82$ vs.\ $-250$~Hz), but English /\textturnr/ F3 retains only ${\sim}5\%$ ($-37$~Hz) and Arabic emphatic F2 ${\sim}4\%$ ($-10$~Hz), the one cell we grade absent. The channel survives where the constriction is a tongue-body gesture the singer must make anyway, not where it is a secondary articulation.

\noindent\textbf{Pitch and voice quality attenuated ---}
The Korean laryngeal F0 contrast retains ${\sim}30\%$ ($+7.7$~Hz vs.\ $+20$--$30$~Hz) rather than vanishing, and the lax--tense H1--H2 contrast ${\sim}19\%$ ($-1.5$ vs.\ ${\sim}8$~dB), while Mandarin tone contours stay near the floor ($+1.6$~Hz). The pattern is competition for the same articulators: the tongue body is not the melody's to use, while duration and F0 are, and give way where it makes a demand.

\subsection{Song-Level Distributions}
\begin{figure}[t]
  \centering
  \includegraphics[alt={Two box-plot panels over the same nine languages, ordered by median nPVI: Arabic, Hindi, Mandarin, Korean, Swedish, Japanese, English, Turkish, French. Left, nPVI falls from about 51 for Arabic to 28 for French. Right, melisma rate is highest for Arabic at 0.11 and lowest for Korean and Swedish near 0.03, so the two rankings do not track each other},width=\linewidth]{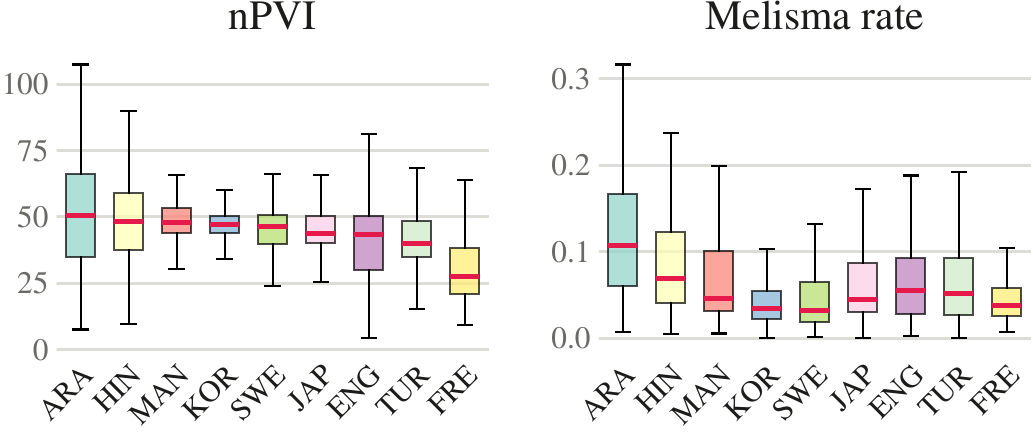}
  \caption{Song-level nPVI (left) and melisma rate (right), 5{,}024 gated
  tracks, 370--751 per language. Boxes span the IQR, median in red, whiskers 1.5\,IQR, outliers omitted. Kruskal--Wallis
  $H = 737$ and $533$ on 8 df; language accounts for
  $\varepsilon^2 = 0.15$ and $0.11$ of the rank variance.}
  \label{fig:song_level}
\end{figure}

Figure~\ref{fig:song_level} shows nPVI and melisma rate over forced-aligned phoneme durations rather than notated values. Both separate the nine languages far beyond chance (Kruskal--Wallis $H = 737$ and $533$ on 8 df, $\varepsilon^2 = 0.15$ and $0.11$). Sub-segmental timing therefore carries language information that notation-based measures do not access. It is not the rhythm-class dimension: the ordering does not track the stress/syllable/mora split, so this is separation rather than a recovery of the speech--music nPVI correlation.

\subsection{Language Classification}
\begin{figure}[t]
  \setlength{\abovecaptionskip}{5pt}
  \centering
  \includegraphics[alt={Nine-by-nine confusion matrix of true against predicted language.
  The diagonal dominates: Arabic 95\%, Turkish 94\%, French 90\%, English 87\%,
  Korean 84\%, Mandarin 83\%, Swedish 82\%, Hindi 81\%, Japanese 73\%. The largest
  off-diagonal cells are Korean read as Mandarin at 10\%, Japanese as Korean at
  9\% and English as Swedish at 9\%},width=\linewidth]{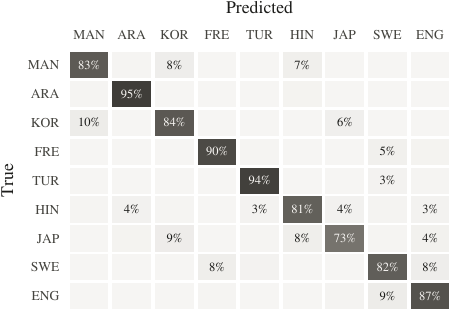}
  \caption{Nine-way language classification from musical prosodic features.
Cells are row-normalized recall, omitted below 3\%. Folds are grouped by
artist.}
  \label{fig:confusion}
  \vspace{-4pt}
\end{figure}

Figure~\ref{fig:confusion} shows the confusion matrix for a Random Forest on 35-dimensional song-level feature vectors, achieving $85.5\%$ balanced accuracy (chance $= 11.1\%$) with artist-grouped folds. Arabic ($95\%$) and Turkish ($94\%$) are the most identifiable and Japanese the least at $73\%$; confusions are structured, not diffuse, pairing Swedish with English and Mandarin with Korean. Grouping costs $1.2$ points against a song-level split ($86.7\%$), bounding what comes from performer identity. The block ablation locates the separability in the marker--control contrasts.

Neither marker rate nor effect magnitude predicts per-language recall (Spearman $\rho = +0.27$ and $+0.50$, $p = 0.49$ and $0.17$), so H2's aggregation prediction is unsupported at $n = 9$. This does not contradict the block ablation. The classifier reads a song's whole profile of contrasts---which dimensions move, in which direction, and how consistently within the song---while the rank test asks only whether languages with larger average effects are easier to identify. A language can separate on a distinctive combination of small contrasts with no large one, and a large effect buys nothing if a competing language carries one of the same size in the same dimension: recall depends on how far a language sits from the others relative to its own spread, not on magnitude against a within-song control. H2 predicts the second drives the first, and nine languages have little power to see it---the effect-magnitude correlation carries the sign H2 requires, but at $n = 9$ that is not separable from noise. We therefore read the prediction as unsupported rather than refuted, as with the Hindi nulls of Section~4.2.
Vocals are source-separated, so residual accompaniment remains an uncontrolled cue.

\begin{table*}[t]
  \centering
  \small
  \setlength{\tabcolsep}{6pt}
  \begin{tabular}{@{}p{1.65cm}llrrrrll@{}}
    \toprule
    \textbf{Category} & \textbf{Language} & \textbf{Feature} & \textbf{Pairs} & \textbf{Speech} & \textbf{Singing} & \textbf{Ratio} & \textbf{Verdict} \\
    \midrule
    \multirow{6}{1.65cm}{\raggedright{Spectral / articulatory}}
      & Mandarin (A) & Retroflex F2 shift \cite{li2008phonetic}
        & 25{,}688
        & ${\sim}$500 Hz &  $+$543 Hz
        &  $\geq$109\% &  \cellcolor[HTML]{2F6DA4}\textcolor[HTML]{FFFFFF}{\emph{preserved}} \\
      & Mandarin (A) & Retroflex centroid \cite{lee2014revisiting}
        & 25{,}899
        & ${\sim}$3{,}000 Hz &  $+$1{,}342 Hz
        &  ${\sim}$45\% & \cellcolor[HTML]{A1BDD6}\textcolor[HTML]{15537E}{\emph{attenuated}} \\
      & Hindi & Retroflex F3 lowering \cite{ladefoged1983non}
        & 4{,}347
        & $-$250 Hz &  $-$82 Hz
        &  ${\sim}$33\% &  \cellcolor[HTML]{BACFE1}\textcolor[HTML]{15537E}{\emph{attenuated}} \\
      & English & /\textturnr/ adj.\ vowel F3 \cite{hagiwara1995acoustic}
        & 78{,}094
        & $-$700 Hz &  $-$37 Hz
        &  ${\sim}$5\% &  \cellcolor[HTML]{F5F8FA}\textcolor[HTML]{15537E}{\emph{attenuated}} \\
      & Arabic & Emphatic F2 lowering \cite{jongman2011acoustics}
        & 25{,}215
        & $-$250 Hz &  $-$10 Hz
        &  ${\sim}$4\% & \cellcolor[HTML]{F7F9FB}\textcolor[HTML]{6B6A66}{\emph{absent}} \\
      & Mandarin (T) & Tone adj.\ vowel F2
        & 152{,}622
        & --- &  $-$204 Hz & --- & \vnov \\
    \midrule
    \multirow{3}{1.65cm}{\raggedright{Voice quality}}
      & Korean & Lax--tense H1--H2 \cite{cho2002acoustic}
        & 36{,}985
        & ${\sim}$8 dB & $-$1.5 dB
        & ${\sim}$19\% & \cellcolor[HTML]{D7E3EE}\textcolor[HTML]{15537E}{\emph{attenuated}} \\
      & Mandarin (A) & Retroflex adj.\ H1--H2
        & 22{,}161
        & --- & $-$0.3 dB & --- & \vnov \\
      & Mandarin (T) & Tone adj.\ H1--H2
        & 142{,}166
        & --- & $+$0.4 dB & --- & \vnov \\
    \midrule
    \multirow{5}{1.65cm}{\raggedright{Temporal / rhythmic}}
      & Mandarin (T) & Tone syllable dur.\ \cite{duanmu2007phonology}
        & 153{,}849
        & ${\sim}$50--120 ms &  $+$26 ms
        &  ${\sim}$30\%\makebox[0pt][l]{\smash{$^{\dagger}$}} &  \cellcolor[HTML]{C1D3E4}\textcolor[HTML]{15537E}{\emph{attenuated}} \\
      & French & Nasal vowel duration \cite{delattre1968role}
        & 34{,}020
        & $+$40 ms &  $+$7.4 ms
        &  ${\sim}$19\%\makebox[0pt][l]{\smash{$^{\dagger}$}} & \cellcolor[HTML]{D7E3EE}\textcolor[HTML]{15537E}{\emph{attenuated}} \\
      & Korean & Asp/tense vs lax dur.\ \cite{cho2002acoustic}
        & 48{,}492
        & ${\sim}$25--40 ms &  $-$10 ms
        &  ---\makebox[0pt][l]{\smash{$^{\dagger}$}} &  \cellcolor[HTML]{CF8C89}\textcolor[HTML]{8B2320}{\emph{reversed}} \\
      & Japanese & Geminate/singleton dur.\ \cite{han1994acoustic}
        & 15{,}213
        & $+$102--204 ms &  $-$15 ms
        & ---\makebox[0pt][l]{\smash{$^{\dagger}$}} & \cellcolor[HTML]{CF8C89}\textcolor[HTML]{8B2320}{\emph{reversed}} \\
      & Arabic & Pharyngeal duration
        & 25{,}498
        & --- &  $+$4 ms & --- & \vnov \\
    \midrule
    \multirow{2}{1.65cm}{\raggedright{Pitch (F0)}}
      & Korean & Asp/tense vs lax F0 \cite{kang2014voice}
        & 25{,}985
        & $+$20--30 Hz &  $+$7.7 Hz
        &  ${\sim}$30\% &  \cellcolor[HTML]{C1D3E4}\textcolor[HTML]{15537E}{\emph{attenuated}} \\
      & Mandarin (T) & Tone contour \cite{schellenberg2012does}
        & 131{,}759
        & --- &  $+$1.6 Hz
        & --- & \vnov \\
    \bottomrule
  \end{tabular}
 \caption{Speech-to-singing preservation ratios: effect sizes documented in related works against our singing measurement.
  \vkey{presbg}{presink}{preserved} $\geq$50\% of the speech effect,
  \vkey{attbg}{attink}{attenuated} 5--50\%,
  \vkey{absbg}{absink}{absent} $<$5\%,
  \vkey{revbg}{revink}{reversed} opposite in sign to speech, so no ratio is
  defined, \textcolor{novink}{\emph{novel}} no baseline and so off the scale. Speech and singing come from different corpora and speakers, so thresholds are conventions. $\dagger$: the aligner compresses duration contrasts, recovering $34\%$ of a long-versus-short difference (Section~\ref{sec:validation}), so temporal ratios are lower bounds; the Japanese baseline converts a $2.0$--$3.0\times$ closure ratio at our control mean of $102$~ms; the tone-contour row has no hertz baseline, and $+1.6$~Hz is $0.6\%$ of the mean F0 of those syllables.}
  \label{tab:preservation}
\end{table*}

\section{Conclusion}\label{sec:conclusion}
\noindent\textbf{Main findings ---}
Three results. First, singing keeps the articulatory gestures the tongue body
must make regardless of the melody and attenuates the rest: the Mandarin
retroflex F2 shift is at least as large in singing as in speech
and retroflex timbre reaches ${\sim}45\%$, while no
durational contrast reaches half its speech baseline (the largest, Mandarin
tonal lengthening, is ${\sim}30\%$, and the aligner compresses these), and the
Korean and Japanese ones reverse sign. Voice quality is attenuated, not eliminated.
Second, the dominant channel is voice quality and spectral tilt
rather than formant transitions: phoneme identity in singing is carried by
source-spectrum properties more than by vocal tract resonance. Third, song-level vectors built from these contrasts support 85.5\% nine-way classification with artist-grouped folds, so the result reflects the language rather than its performers. Since neither marker
frequency nor effect magnitude predicts per-language accuracy at $n = 9$, this
establishes separability, not the accumulation H2 predicts.

\noindent\textbf{Discussion ---}
Within-song matched pairs hold singer, genre, and melody constant, so H1
effects are consequences of vocal tract physics rather than musical tradition.
At the language level (H2) no such separation is possible---a dark Arabic
timbre may reflect cumulative pharyngeal articulation, a tradition that evolved around it, or both. Distinguishing them needs variation that phonology and tradition do not share, which chart data does not supply. We therefore treat H1 as established and H2 as empirically demonstrated but causally open. Our vocals are separated, not natively unaccompanied. For MIR this makes language an audible property of a vocal track, and one that sits in phonation and spectral shape rather than the pitch and timing singing-voice models are usually evaluated on.

\noindent\textbf{Limitations and future work ---}
Nine languages exclude click languages, ejective-rich inventories, and richer tone systems. Our 2018--24 chart corpus ties
each language to one or two countries. Phoneme boundary errors propagate to
every feature and set a floor on what the design can measure: single-phone
effects are recoverable only when onset error is well below phone duration. Preservation ratios use
published speech baselines from different corpora, speakers,
and protocols, so they are order-of-magnitude comparisons rather
than matched contrasts. A second constraint is lyric
quality---$46\%$ of tracks fail the agreement gate, concentrated in Arabic and Swedish, needing better lyrics, not a better
aligner. Residual accompaniment survives separation in the timbral and spectral channels, where our largest effects sit. And we measure acoustics, not audibility: a perceptual test is the probe. H3 remains untested; the natural probe is cover songs and multilingual
singers.

\section{Acknowledgments}
This work was partly supported by the Youlchon Foundation (Nongshim Corporation and affiliated companies), Korea [50\%]; by the National Research Foundation of Korea (NRF) under grant [No. RS-2025-24683892, 45\%]; and by the Institute of Information \& communications Technology Planning \& Evaluation (IITP) grant [No. RS-2021-II211343, 5\%], funded by the Korea government (MSIT).

\section{AI Usage Statement}
Generative AI tools were used for code development assistance and language polishing. All research design, experiments, analysis, and conclusions were produced by the authors.


\bibliography{ref}

\end{document}